\documentclass[fleqn,usenatbib]{mnras}

\usepackage{newtxtext,newtxmath}

\usepackage[T1]{fontenc}
\usepackage{ae,aecompl}

\usepackage{graphicx}	
\usepackage{amsmath}	
\usepackage{amssymb}	

\title[The SMC cluster Lindsay 1]{An Extragalactic Chromosome Map: The intermediate age SMC Cluster Lindsay 1}

\author[S. Saracino et al.]{S. Saracino$^{1,2}$\thanks{E-mail: s.saracino@ljmu.ac.uk},
N. Bastian$^{1}$,
V. Kozhurina-Platais$^{3}$,
I. Cabrera-Ziri$^{4}$\thanks{Hubble Fellow},
E. Dalessandro$^{2}$,
\newauthor
N. Kacharov$^{5}$,
C. Lardo$^{6}$,
S. S. Larsen$^{7}$,
A. Mucciarelli$^{2,8}$,
I. Platais$^{9}$,
M. Salaris$^{1}$
\\
$^{1}$Astrophysics Research Institute, Liverpool John Moores University, 146 Brownlow Hill, Liverpool L3 5RF, UK\\
$^{2}$INAF-Osservatorio di Astrofisica \& Scienza dello Spazio, via Gobetti 93/3, I-40129, Bologna, Italy\\
$^{3}$Space Telescope Science Institute, 3700 San Martin Drive, Baltimore, MD 21218, USA\\
$^{4}$Harvard-Smithsonian Center for Astrophysics, 60 Garden Street, Cambridge, MA 02138, USA\\
$^{5}$Max-Planck-Institut f\"ur Astronomie, K\"onigstuhl 17, D-69117 Heidelberg, Germany\\
$^{6}$Laboratoire d'astrophysique, \' Ecole Polytechnique F\' ed\' erale de Lausanne (EPFL), Observatoire, 1290, Versoix, Switzerland\\
$^{7}$Department of Astrophysics/IMAPP, Radboud University, P.O. Box 9010, 6500 GL Nijmegen, The Netherlands\\
$^{8}$Dipartimento di Fisica \& Astronomia, Universit\` a degli Studi di Bologna, via Gobetti 93/2, I-40129, Bologna, Italy\\
$^{9}$Department of Physics and Astronomy, Johns Hopkins University, 3400 North Charles Street, Baltimore, MD 21218, USA\
}

\date{Accepted XXX. Received YYY; in original form ZZZ}

\pubyear{2019}

\begin{document}
\label{firstpage}
\pagerange{\pageref{firstpage}--\pageref{lastpage}}
\maketitle

\begin{abstract}
The discovery of star-to-star abundance variations (a.k.a. multiple populations - MPs) within globular clusters (GCs), which are generally not found in the field or in lower mass open clusters, has led to a search for the unique property of GCs that allow them to host this phenomenon. Recent studies have shown that MPs are not limited to the ancient GCs but are also found in massive clusters with ages down to (at least) 2~Gyr. This finding is important for understanding the physics of the MP phenomenon, as these young clusters can provide much stronger constraints (e.g. on potential age spreads within the clusters) than older ones. However, a direct comparison between ancient GCs and intermediate clusters has not yet been possible due to the different filters adopted in their studies. 
Here we present new HST UV photometry of the 7.5~Gyr, massive SMC cluster, Lindsay 1, in order to compare its pseudo colour-colour diagram to that of Galactic GCs. We find that they are almost identical and conclude that the MPs phenomenon is the same, regardless of cluster age and host galaxy.
\end{abstract}



\section{Introduction}
For many years globular clusters (GCs) have been generally considered the best example of Simple Stellar Populations (i.e. collections of stars having same age and metallicity). However, the detection of star-to-star abundance spreads in light-elements (e.g. C, N, O, Na) through photometric and spectroscopic observations (see \citealt{gratton2012} and \citealt{bastian2018}) has totally revolutionized this view, leading to a systematic search of the main drivers of the multiple populations (MPs) phenomenon in GCs. Many scenarios have been proposed over the years (e.g. \citealt{decressin2007,dercole2008,bastian2013,denissenkov2014,gieles2018}) but a scenario that can self-consistently explain all observational findings is still missing. In fact, recent studies have shown that MPs are not limited to the ancient Galactic GCs \citep{piotto2015,milone2017} but are also found in extragalactic massive clusters (e.g. Magellanic Clouds (LMC/SMC), \citealt{mucciarelli2009,dalessandro2016}; M81, \citealt{mayya2013}; Fornax, \citealt{larsen2012,larsen2018}; M31, \citealt{nardiello2019,schiavon2013}) with ages down to (at least) 2 Gyr \citep{niederhofer2017a,niederhofer2017b,martocchia2018}.\\
One of the main questions that it is now time to address is whether what we see is the manifestation of the same phenomenon in different environments/ages or instead we need to invoke different formation mechanisms for MPs in different galaxies. In the last years, two Hubble Space Telescope (HST) surveys of Milky Way (MW) GCs \citep{piotto2015,nardiello2018} and LMC/SMC stellar clusters \citep{niederhofer2017a,niederhofer2017b,martocchia2017,martocchia2018} have been devoted to the search of MPs but their strategies (in particular the filter combinations adopted) did not allow for a direct comparison of the results. To overcome this issue, we undertook an ultraviolet (UV) HST survey of five LMC/SMC massive clusters with ages ranging from 1.7 Gyr to 8 Gyr, thus we now have all the ingredients to put stellar clusters from different galaxies, and at different ages, in the same framework. In particular, we can use the ``chromosome map" \citep{milone2017}, a pseudo colour diagram that is extremely efficient at finding and quantifying the presence of MPs in clusters.

Here we present the results obtained for a cluster in our sample: Lindsay 1, an intermediate age cluster (7.5 Gyr, \citealt{glatt2008}) in the SMC, with a mass of $\sim2 \times 10^{5} M_{\odot}$ \citep{mclaughlin2005} and a metallicity of [Fe/H] $\sim -1.28$ (Z=0.001, \citealt{niederhofer2017b}). A spectroscopic study by \citet{hollyhead2017} found a significant N-spread among 37 RGB stars in the cluster, which was photometrically confirmed with a larger sample of stars using HST observations \citep{niederhofer2017b}.

This paper is structured as follows. In Section \ref{sec:obs} the observational database, the photometric reduction and the analysis are presented. In Section \ref{sec:res} we discuss our results, based on the chromosome map as a diagnostic for MPs, then comparing these findings against the benchmarks of MW GCs. In Section \ref{sec:concl} we summarise the most relevant results and draw our conclusions. 
\vspace{-0.4cm}
\section{Observations and data reduction}\label{sec:obs}
\subsection{Data-sets and photometry}
\begin{figure}
    \centering
	\includegraphics[width=0.43\textwidth]{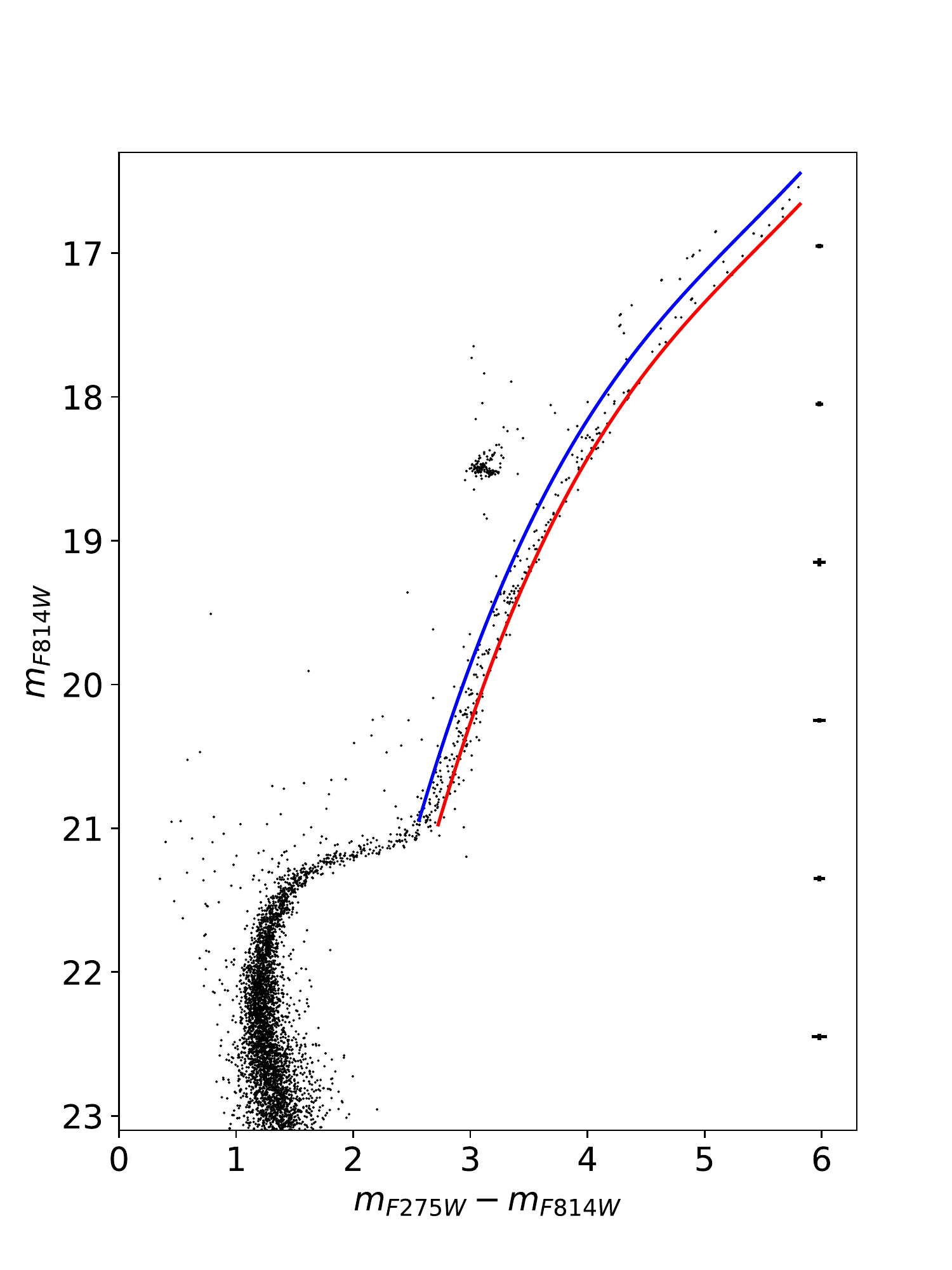}
    \caption{($m_{F814W}$, $m_{F275W}-m_{F814W}$) CMD of all the stars in common between the ACS and WFC3 catalogues used in this work. Blue and red lines represent the adopted fiducial lines in the analysis (see Section \ref{sec:cmap} for details). The photometric errors for each bin of $m_{F814W}$ magnitudes are shown on the right side of the panel.}
    \label{fig:cmd}
\end{figure}

This work is based on three different sets of HST observations. One consists of archival observations acquired with the optical F555W filter (two with $t_{exp}$=20s and four with $t_{exp}$=496s) and F814W filters (two with $t_{exp}$=10s and four with $t_{exp}$=474s) from the program GO-10396 (PI: J. Gallagher), using the Advanced Camera for Surveys (ACS). The second program is composed of near-UV and optical images obtained through the Wide Field Camera 3 (WFC3) UVIS channel under the LMC/SMC survey of GCs (GO-14069, PI: N. Bastian). Three images were acquired in each of F336W and F438W filters with a total exposure time of 2900s and 1040s, respectively. Along with these data, already used by \citet{niederhofer2017b} to investigate the presence of MPs in Linsday 1, here we added new UV observations recently acquired with the F275W filter under the ongoing HST program (GO-15630, PI: N. Bastian): six images for a total $t_{exp}$ of 9000s. These observations are only 1/3 of the total number of images we can count on at the end of the survey. However, they are still good enough to put Linsday 1 on the same footing of MW GCs hosting MPs. In the three data-sets, an appropriate dither pattern of a few arcseconds has been adopted for each pointing in order to fill the inter-chip gaps and avoid spurious effects such as bad pixels and cosmic rayes. 
The photometric analysis has been performed on images processed, flat-fielded, bias subtracted, and corrected for Charge Transfer Efficiency losses by standard HST pipelines ($\_{flc}$ images). We derived the stellar photometry using the spatially variable `effective point spread function' (ePSF) method \citep{anderson2006} for both WFC3/UVIS and ACS/WFC. Instrumental magnitudes of the catalogues were calibrated to the VEGAMAG photometric system, applying the zero points of ACS/WFC and WFC3/UVIS from the instruments web-site respectively.
The corrected positions for geometric distortions \citep{bellini2011} were then transformed to the absolute coordinate system (RA, Dec) by using the stars in common with the Gaia Data Release 2 (DR2, \citealp{gaia2016,gaia2018}) and by means of the cross-correlation software CataXcorr \citep{montegriffo1995}. 
The resulting ($m_{F814W}$, $m_{F275W}-m_{F814W}$) colour-magnitude diagram (CMD) of Lindsay 1 is shown in Figure~\ref{fig:cmd}. 
\subsection{Differential Reddening}
Inspecting the CMD in Figure \ref{fig:cmd}, it is quite evident that Lindsay 1 is not severely affected by differential reddening across the FOV exploited in this study: the evolutionary sequences are very well defined. However, in order to be consistent with all the previous works, we corrected for the effect of the differential extinction, using the approach described in \citet[][see also \citealt{dalessandro2018}]{saracino2019}. Briefly, first we created the cluster mean ridge line (MRL) in the ($m_{F814W}$, $m_{F555W}-m_{F814W}$) CMD, then we selected a sample of bona-fide stars in the magnitude range 19.5$<m_{F814W}<$23.5 and we computed the geometrical distance ($\Delta X$) of those stars from the MRL. This reference sample has been then used to assign a $\Delta X$ value to each star in our photometric catalogue, by looking at its 25 closest reference stars. Using the extinction coefficients from \citet{cardelli1989}, we finally transformed $\Delta X$ of each star into the local differential reddening $\delta E(B-V)$. The resulting $\delta E(B-V)$ are very low (a mean value of 0.003 and a maximum variation of about 0.01 in the FOV) and comparable with the photometric uncertainties in the optical filters (shown in Figure 1, right side), thus not producing any significant difference in the cluster CMD.

Lindsay 1 is a well extended cluster (core radius $r_{c}$ = 61.7 arcsec, \citealt{glatt2009}) so that a statistical background subtraction within the field of view of our observations cannot be performed. However, as suggested by \citet{niederhofer2017b} (see also \citealt{parisi2016}), Lindsay 1 is expected to be only slightly affected by field star contaminants, due to its location in the outskirts of the SMC. Nevertheless, we tried to look for star interlopers by using Gaia DR2 \citep{gaia2018} proper motions. As expected, due to the cluster distance, we found only a few bright stars in common with our photometry. Unfortunately, most of these DR2 counterparts have too large proper motion errors to be meaningful in the context of SMC clusters (see \citealt{platais2018} for further details). However, even if we could not clean our sample for field star contaminants, we do not expect they may have an impact on our conclusions.
\section{Results}\label{sec:res}
\subsection{Chromosome Map}\label{sec:cmap}
\begin{figure*}
	\includegraphics[width=0.95\textwidth]{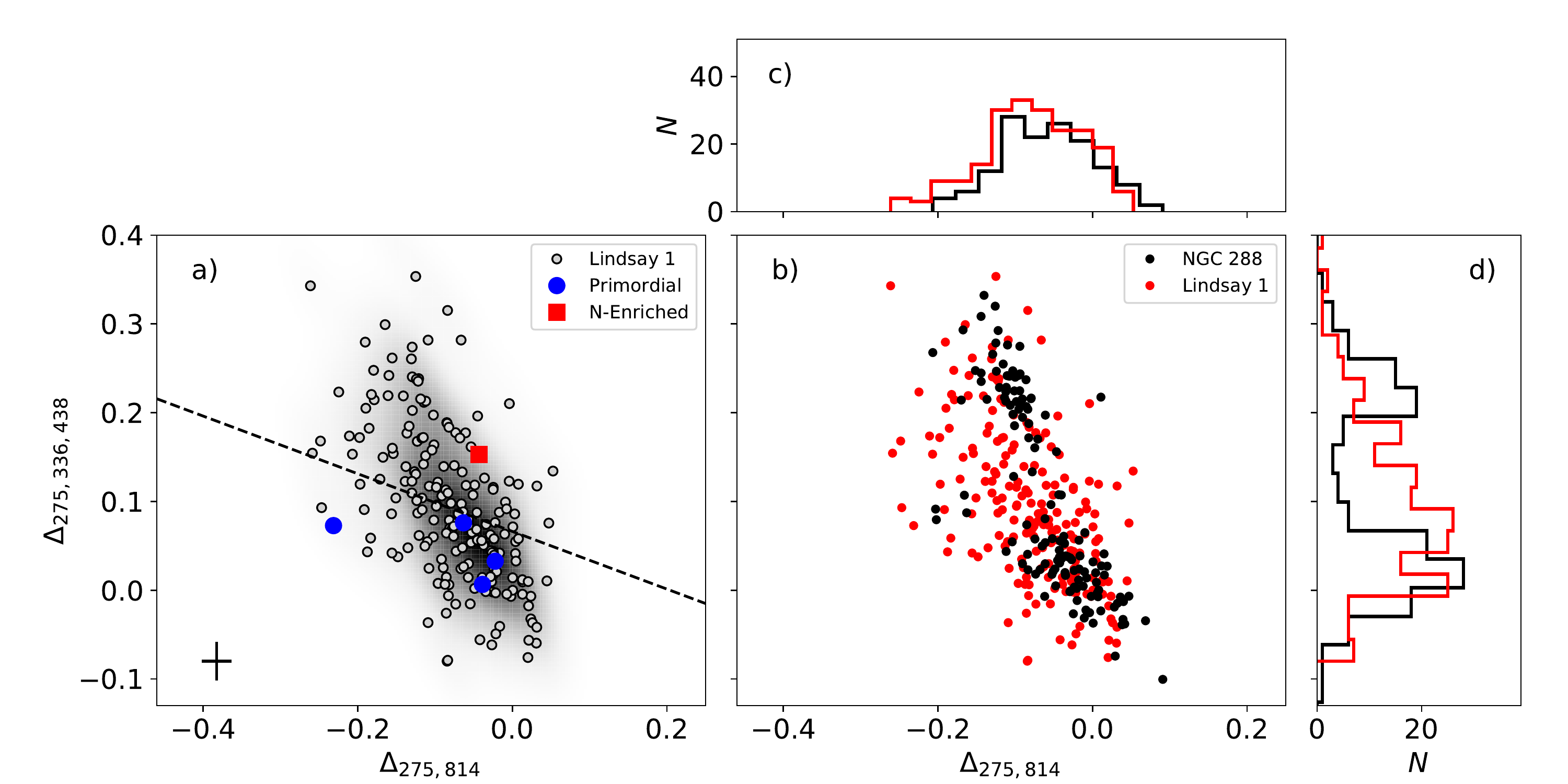}
    \caption{{\it Panel a}: ($\Delta_{275,814}$, $\Delta_{275,336,438}$) colour-colour diagram (``chromosome map") of Lindsay 1 (light-grey points). The related KDE distribution is shown in greys-scale. The stars in common with the sample of \citet{hollyhead2017} are marked as blue circles (primordial composition) and a red square (enriched). For reference, the separation between normal and enriched stars in NGC 288 (from \citealt{milone2017}) is superimposed as a black dashed line. {\it Panel b}: The chromosome map of the cluster (red points) is compared to that of NGC 288 (black points). {\it Panel c}: The histogram of the verticalised colour distribution $\Delta_{275,814}$ of RGB stars of Lindsay 1 is shown, together with that of NGC 288. {\it Panel d}: Same as in panel {\it c} but for the verticalised pseudo colour distribution $\Delta_{275,336,438}$ of the same RGB stars in the clusters. The colour codes in panels {\it b}, {\it c} and {\it d} are the same.}
    \label{fig:chromo}
\end{figure*}

In \citet{niederhofer2017b}, the authors reported the detection of MPs in the CMD of Lindsay 1, using the filter combination of the form (F336W-F438W)-(F438W-F343N) = $C_{F336W,F438W,F343N}$ as a good diagnostic to separate populations with different C and N abundances. However, the MPs phenomenon has been recently comprehensively investigated via the chromosome map \citep{milone2015,milone2017}, a pseudo colour-colour plot using F275W, F336W, F438W and F814W filters. In order to further test the presence of MPs in Lindsay 1, we produced the chromosome map of the cluster, using the procedure outlined in \citet{milone2017}. We first used the ($m_{F814W}$, $m_{F555W}-m_{F814W}$) CMD to select bona-fide RGB stars in the magnitude range 16.5$<m_{F814W}<$21. Then we used the ($m_{F814W}$, $m_{F275W}-m_{F814W}$) CMD to define two fiducial lines (Figure \ref{fig:cmd}) as the 10th and 90th percentiles of the $m_{F275W}-m_{F814W}$ distribution of the previously selected RGB stars. We then verticalised the distribution of RGB stars and normalized them for the intrinsic RGB width at 2 mag brighter than the turn-off, thus creating the $\Delta_{F275W,F814W}$. 
The histogram of $\Delta_{F275W,F814W}$ is shown by red line in panel {\it c} of Figure \ref{fig:chromo}. We applied the same approach to the pseudo colour diagram ($m_{F814W}$, $C_{F275W,F336W,F438W}$\footnote{$C_{F275W,F336W,F438W}$ = (F275W-F336W)-(F336W-F438W) as defined in \citet{milone2017}.}), in order to derive $\Delta_{F275W,F336W,F438W}$. The histogram of $\Delta_{F275W,F336W,F438W}$ is reported in panel {\it d}, red line.

These values have been used to compute the ($\Delta_{F275W,F814W}$, $\Delta_{F275W,F336W,F438W}$) chromosome map of Lindsay 1 presented in panel {\it a} of Figure \ref{fig:chromo} as light-grey points and to evaluate the kernel density distribution (KDE) from a Gaussian kernel for both axis, shown in greys-scale in the figure. The observed distribution of stars is wider than what expected from photometric errors alone (shown in panel {\it a}, bottom-left side). It also shows the same shape observed in Galactic GCs that host MPs: a further evidence of the presence of MPs in this intermediate-age cluster.

From \citet{milone2017}, we expect N-normal (first population - FP) stars to be located around ($\Delta_{275,814}$, $\Delta_{275,336,438}$) $\sim$ (0,0), while N-enriched (second population - SP) stars at larger values of $\Delta_{275,336,438}$. This is confirmed by the following consistency check. We cross-correlated the spectroscopic results by \citet[][]{hollyhead2017} with our photometric data and we found five stars in common. One of them was determined to be enhanced in N, while the other four have a primordial composition. They are shown as red square and blue circles, respectively, on top of the chromosome map of Lindsay 1 in panel {\it a} of Figure \ref{fig:chromo}.
\subsection{First vs Second Population}
\begin{figure}
    \centering
	\includegraphics[width=0.43\textwidth]{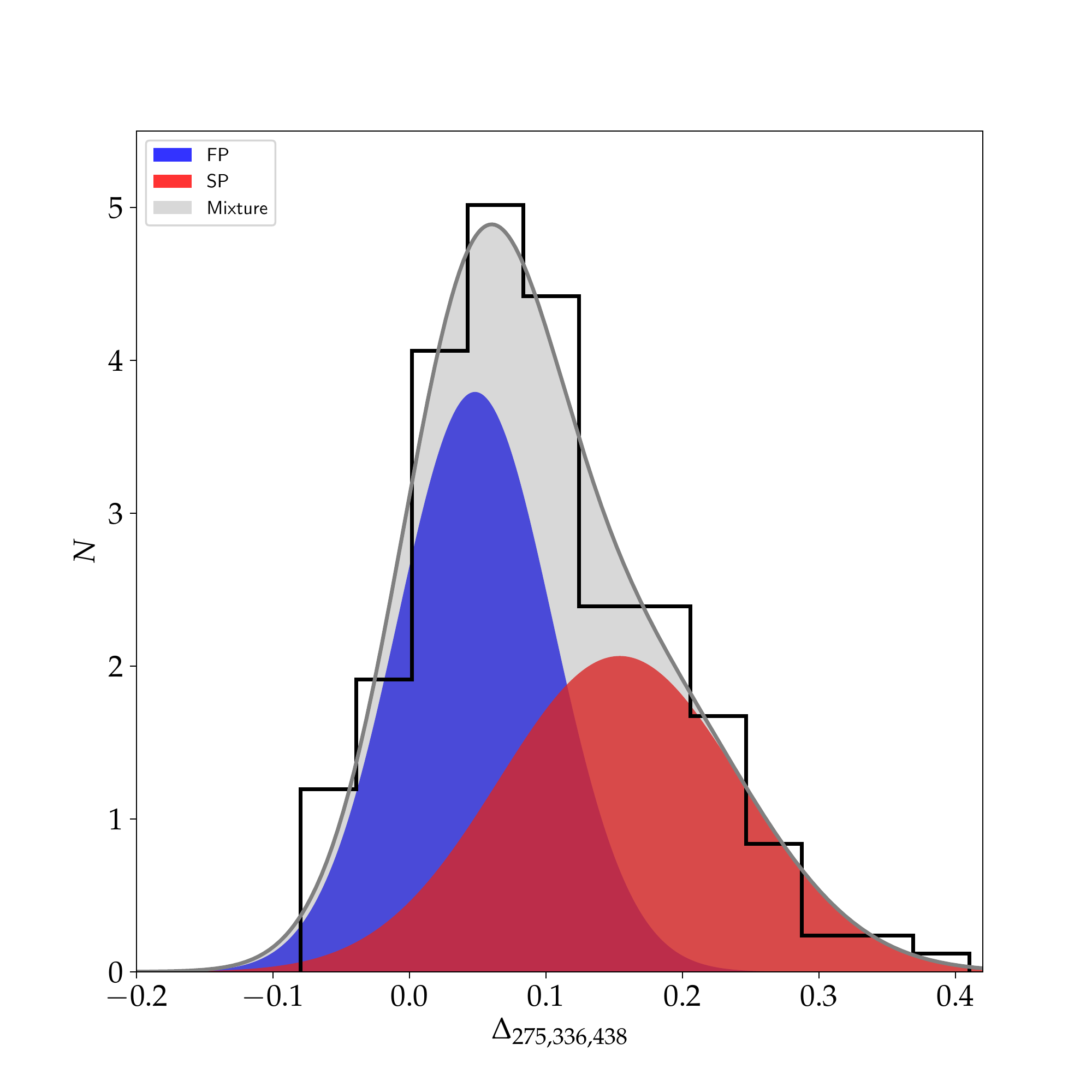}
    \caption{Histogram of the verticalized pseudo colour distribution $\Delta_{275,336,438}$ of RGB stars in Lindsay 1. The Gaussian Mixture Model that fit best the unbinned data is overimposed, as a grey shaded area and a solid line. The individual Gaussians are instead shown as blue and red regions, representing the predicted first and second population of stars, respectively.}
    \label{fig:gauss}
\end{figure}

As already noted, RGB stars of Lindsay 1 show an extended pseudo colour $\Delta_{F275W,F336W,F438W}$ distribution (see panel {\it d} of Figure \ref{fig:chromo}), with a slight hint of bimodality. In order to infer whether the distribution can be fitted by single or multiple gaussians, we applied the Gaussian Mixture Models analysis \citep{muratov2010} on our unbinned sample of $\Delta_{F275W,F336W,F438W}$. The result is shown in Figure \ref{fig:gauss} and it demonstrates the data are best fitted with a two-component gaussian (grey shaded area), as the sum of two gaussians having peaks at $\Delta_{F275W,F336W,F438W}$ $\sim$ 0.041 and $\sim$ 0.154, respectively (blue and red shaded areas). 
We classified the two sub-populations as FP and SP moving from blue to red colours and from the areas under the Gaussian functions we computed the number ratios between the sub-populations. We find that 111 stars can be attributed to the FP and 94 to the SP population, thus yielding $N_{FP}/N_{TOT}$ = 0.54$\pm$0.09 and $N_{SP}/N_{TOT}$ = 0.46$\pm$0.09, where $N_{TOT}$ is the total number of the selected sample of RGB stars\footnote{We verified that the $N_{FP}/N_{TOT}$ and $N_{SP}/N_{TOT}$ vary only by 3\% in favor of the FP, when a counterclockwise rotation of $18^{\circ}$ in $\Delta_{F275W,F336W,F438W}$ is considered \citep{milone2017}.}. The errors on both fractions have been computed by considering two terms: a systematic error related to the adopted fiducial lines ($\sim$0.07), and the poissonian error ($\sim$0.05). The fraction of enriched stars found here is somewhat higher compared to the 36\% found by \citet[][]{niederhofer2017b}, but both results can be brought to agreement, once the differences in the RGB fiducial lines as well as in the adopted pseudo color distributions, are properly taken into account. 
\subsection{Comparison with MW GCs}\label{sec:concl}
An interesting step forward in the understanding of the MPs phenomenon comes from the comparison of the behaviour of the intermediate-age cluster Lindsay 1 against the benchmarks of the survey results of old Galactic GC \citep{piotto2015,milone2017,nardiello2018}, in order to see whether they share the same properties. To do so, in panel {\it b} of Figure \ref{fig:chromo} we compared the chromosome map of Lindsay 1 (red points) with that of NGC 288, a MW GC with similar metallicity ([Fe/H]$\sim$ -1.3, \citealt{carretta2009}, black points)\footnote{NGC 288 has been used as a reference, but the chromosome maps of other GCs with similar metallicity (e.g. NGC 1261) have been also compared, reaching the same conclusions.}. The histograms of the $\Delta_{F275W,F814W}$ and $\Delta_{F275W,F336W,F438W}$ distributions have been
also presented in panels {\it c} and {\it d}. 
The SMC cluster Lindsay 1 shows exactly the same pattern as NGC 288, with the same orientation angle and approximately the same extension. 
The dashed black line in panel {\it a} of this figure is taken from \citet[][]{milone2017}, and separates the normal and enriched populations in the comparison cluster NGC 288. We can notice that the enriched star in Lindsay 1 lies above this line, while the other four stars are below the line, consistently with the separation found in NGC 288. Moreover, we know that a spread in $\Delta_{F275W,F336W,F438W}$ reveals the presence of N-enriched stars, while a spread in $\Delta_{F275W,F814W}$ is caused by a range of helium (Y) abundances (\citealt{lardo2018,milone2018}). The presence of a Y-spread in Lindsay 1, as shown by the chromosome map, is consistent with the results by \citet[][]{chantereau2019} who determined $\Delta Y \sim 0.03$, by modeling the horizontal branch (HB) of the cluster. Also \citet{milone2018} determined $\Delta Y \sim 0.02$ for NGC 288 from its RGB colours, that is consistent with \citet[][]{chantereau2019} estimate from the HB.
These comparisons strongly suggest that MPs in an intermediate age massive cluster like Lindsay 1 have a common origin with the MPs observed in old Galactic GCs.

\begin{figure}
    \centering
	\includegraphics[width=0.43\textwidth]{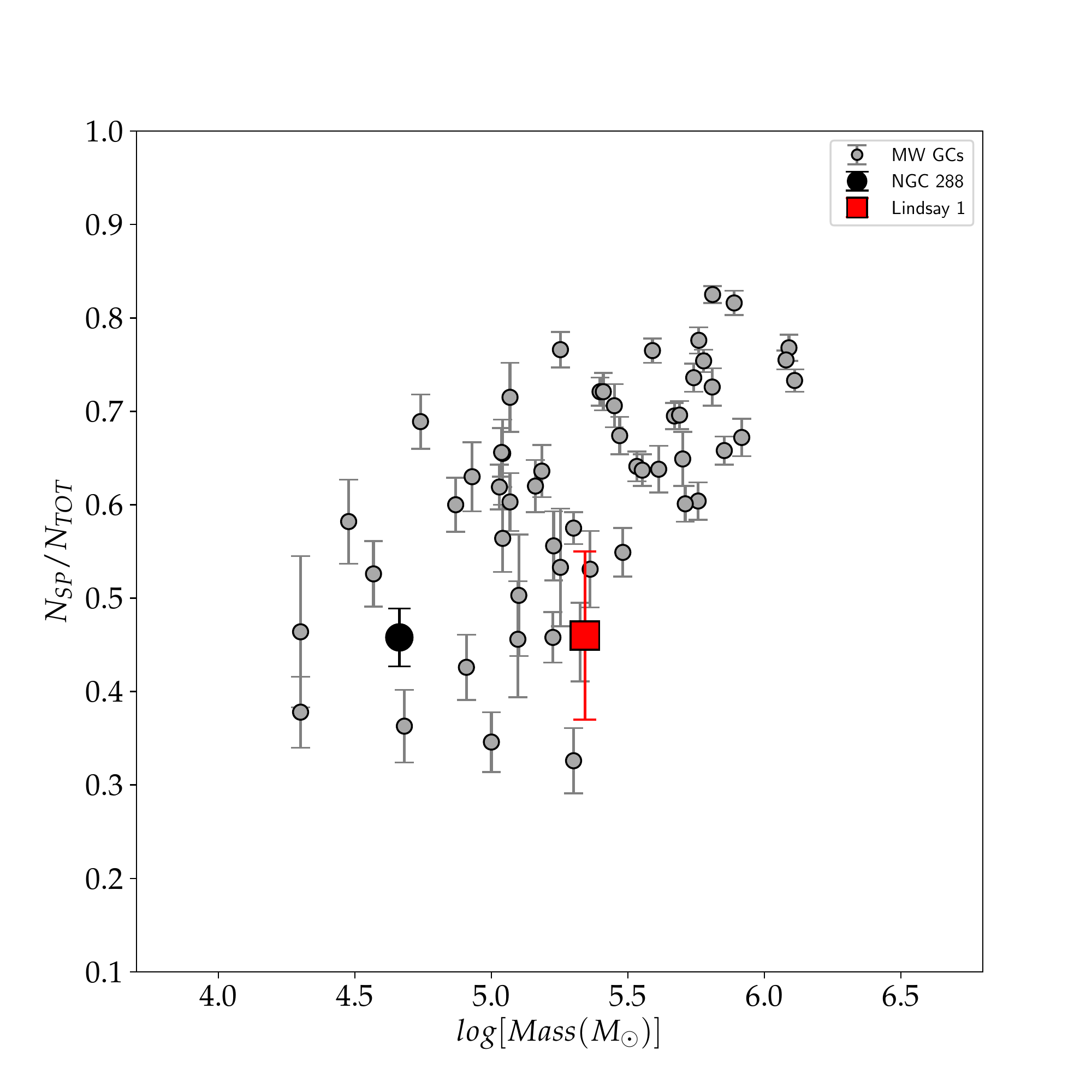}
    \caption{Fraction of SP stars relative to the total number of stars, as measured on the RGB, as a function of the cluster mass. Gray points represent MW GCs observed within the HST UV Legacy Survey, whereas Lindsay 1 is highlighted as a red square. The black dot instead refers to NGC 288, for comparison.}
    \label{fig:nSP}
\end{figure}

Finally, we compared the fraction of SP stars in this cluster, with the trend found in MW GCs. As shown in Figure \ref{fig:nSP}, the derived value of $N_{SP}/N_{TOT}$ for Lindsay 1 appears to be compatible with the general ($N_{SP}/N_{TOT}$, Mass) trend observed for the clusters of the HST UV Legacy Survey \citep{piotto2015,nardiello2018} analysed with the same observational strategy (instrument and filters) as for Lindsay 1. At the same time, the evidence that the fraction of SP stars in this cluster is slightly lower than what found in MW GCs with similar mass is consistent with the results of NGC 121, another SMC cluster ($N_{SP}/N_{TOT}$ $\sim$ 0.30 - 0.35, \citealt{niederhofer2017a}, \citealt{dalessandro2016}).
\section{Discussion and Conclusions}
\label{sec:concl} 
In this study we complemented high-resolution near-UV/optical archival HST images of the SMC cluster Lindsay 1, with new HST observations recently acquired in the F275W filter. These new data gave us the opportunity to further investigate the presence of MPs in the cluster, by exploiting the power of the ``chromosome map" \citep{milone2017}, a pseudo colour-colour plot using the HST F275W, F336W, F438W and F814W filters, to separate sub-populations of stars having different chemical abundances. It is the first time that this has been done for a cluster younger than 10 Gyr, or belonging to a (non-accreted) galaxy outside the MW.
The UV-optical photometry, corrected for differential extinction, revealed a slight broadening of the RGB of Lindsay 1, also detected in the chromosome map ($\Delta_{F275W,F814W}$, $\Delta_{F275W,F336W,F438W}$ diagram), where RGB stars extend beyond their errors in the expected direction, thus confirming Lindsay 1 hosts stars with different light-element abundances. This is in agreement with the previous results on the cluster, from both a photometric \citep{niederhofer2017b} and a spectroscopic point of view \citep{hollyhead2017}. The stars in common with the spectroscopic targets of \citet{hollyhead2017} are also exactly located where N-normal and N-enriched stars are expected to be in the chromosome map. 
From the pseudo colour $\Delta_{F275W,F336W,F438W}$ we found that $\sim54\%$ of RGB stars belong to the FP, while the remaining $\sim 
46\%$ is part of the SP. These percentages demonstrate that the pseudo colour $C_{F336W,F438W,F343N}$ adopted in recent papers to detect MPs in LMC/SMC clusters \citep{niederhofer2017a,niederhofer2017b,martocchia2017,martocchia2018} did a good job in separating populations with light-elements abundance variations, at least in N.\\
One of the most interesting results of this study comes from the comparison of the chromosome map of Lindsay 1 with that of NGC 288, a Galactic GC with similar metallicity but considerably older ($\sim$12 Gyr, \citealt{dotter2010}). The two clusters occupy the same parameter space, with the same orientation angle. Interestingly, the separation between FP and SP in the two clusters look to be roughly the same. Moreover, the value $\Delta Y \sim 0.03$ determined by \citet{chantereau2019} for Lindsay 1 is consistent with the $\Delta_{F275W,F814W}$ observed distribution, with respect to that of NGC 288 ($\Delta Y \sim 0.02$, \citealt{milone2018}). This is the first clear evidence that the main drivers for the MPs phenomenon are the same in time and space (i.e. for clusters with different age and host galaxy). Moreover, the fraction of SP stars with respect to the total number of RGB stars appears to be compatible with the general ($N_{SP}/N_{TOT}$, Mass) trend observed for MW GCs \citep{milone2017}, although slightly lower. These results are expected to be further improved in the following months, as soon as new F275W observations for Lindsay 1 and four younger clusters, will become available.
\section*{Acknowledgements}
We thank the anonymous referee for the careful reading of the paper, and for the useful comments and suggestions.\\ SS and NB gratefully acknowledge financial support from the European Research Council (ERC-CoG-646928, Multi-Pop). NB also acknowledges support from the Royal Society (University Research Fellowship). V.K-P. is very gratitude to Jay Anderson for sharing with us his ePSF code. CL thanks the Swiss National Science Foundation for supporting this research through the Ambizione grant number PZ00P2 168065. The authors gratefully acknowledge grant support for programs GO-14069, GO-15630, provided by NASA through Hubble Fellowship grant HST-HF2-51387.001-A awarded by the Space Telescope Science Institute, which is operated by the Association of Universities for Research in Astronomy, Inc., for NASA, under contract NAS5-26555.



\bibliographystyle{mnras}
\bibliography{lind1} 
\bsp	
\label{lastpage}
\end{document}